\documentclass[aps,prl,twocolumn,showpacs,preprintnumbers,amsmath,amssymb,superscriptaddress,longbibliography]{revtex4-2}
\usepackage{hyperref}
\hypersetup{colorlinks=true, citecolor=blue, urlcolor=blue, linkcolor=blue,urlcolor=cyan}
\usepackage{color}
\usepackage{graphicx}
\usepackage{dcolumn}
\usepackage{braket}
\usepackage{amsthm}
\usepackage[caption=false]{subfig}

\newtheorem{theorem}{Theorem}

\def\be{\begin{equation}}
\def\ee{\end{equation}}
\def\ba{\begin{eqnarray}}
\def\ea{\end{eqnarray}}

 \def\ba{{\bar{\alpha}}}

\def\CO{{\mathcal{O}}}

\begin{document}

\title{Efficient Simulation of Low Temperature Physics in One-Dimensional Gapless Systems}
\preprint{CALT-TH 2023-018}
\preprint{RIKEN-iTHEMS-Report-23}
\preprint{YITP-23-74}

\author{Yuya Kusuki}
\affiliation{\it Walter Burke Institute for Theoretical Physics, California Institute of Technology, Pasadena, CA 91125, USA}
\affiliation{\it Interdisciplinary Theoretical and Mathematical Sciences (iTHEMS), RIKEN, Wako, Saitama 351-0198, Japan}

\author{Kotaro Tamaoka}
\affiliation{\it Department of Physics, College of Humanities and Sciences, Nihon University, Tokyo 156-8550, Japan}

\author{Zixia Wei}
\affiliation{\it Center for the Fundamental Laws of Nature \& Society of Fellows, Harvard University, Cambridge, MA 02138, USA
}
\affiliation{\it Interdisciplinary Theoretical and Mathematical Sciences (iTHEMS), RIKEN, Wako, Saitama 351-0198, Japan}
\affiliation{\it Yukawa Institute for Theoretical Physics, Kyoto University, Sakyo-ku, Kyoto 606-8502, Japan}

\author{Yasushi Yoneta}
\affiliation{\it Center for Quantum Computing, RIKEN, Wako, Saitama 351-0198, Japan}
\affiliation{\it Department of Basic Science, The University of Tokyo, Meguro, Tokyo 153-8902, Japan}

\begin{abstract}
We discuss the computational efficiency of the finite temperature simulation with the minimally entangled typical thermal states (METTS).
To argue that METTS can be efficiently represented as matrix product states, we present an analytic upper bound for the average entanglement R\'enyi entropy of METTS for R\'enyi index $0<q\leq 1$.
In particular, for 1D gapless systems described by CFTs, the upper bound scales as $\mathcal{O}(c N^0 \log \beta)$ where $c$ is the central charge and $N$ is the system size.
Furthermore, we numerically find that the average R\'enyi entropy exhibits a universal behavior characterized by the central charge and is roughly given by half of the analytic upper bound.
Based on these results, we show that METTS can provide a speedup
compared to employing the purification method to analyze thermal equilibrium states at low temperatures in 1D gapless systems.

\end{abstract}

\maketitle


\textit{Introduction.---}
Simulating quantum many-body systems at finite temperature is an everlasting topic in statistical physics and computational physics. While thermal equilibrium states at inverse temperature $\beta$ of a quantum system with Hamiltonian $H$ can be described by the canonical Gibbs state
\begin{align}
    \rho_\beta^{\rm(can)} \equiv \frac{e^{-\beta H}}{{\rm Tr}\left(e^{-\beta H}\right)},
\end{align}
the computational cost to construct it directly diverges exponentially with system size, which makes it unrealistic in practice to be accurately prepared.

The minimally entangled typical thermal states (METTS) algorithm \cite{White09,SW10} is one of the methods developed to efficiently simulate finite temperature quantum systems. Consider a spin-$1/2$ chain of length $N$, the orthogonal $2^N$ classical product states $\ket{P_i}=\otimes_{\mathrm{sites}\ n} \ket{i_n}$
form a complete basis of the whole Hilbert space, where $\{\ket{i_n}\}$ is an orthogonal basis of each site. METTS are a class of states defined as 
\begin{align}\label{eq:METTS_def}
    \ket{\mu_i} = \frac{e^{-\beta H/2} \ket{P_i}}{\sqrt{\braket{P_i|e^{-\beta H}|P_i}}}.
\end{align}
By construction, $\rho_{\beta}^{\rm (can)}$ can be decomposed as a classical mixture of METTS
\begin{align}
    \rho_{\beta}^{\rm (can)} = \sum_i p_i |\mu_i \rangle\langle \mu_i|,~~
    p_i \equiv \frac{\braket{P_i|e^{-\beta H}|P_i}}{\sum_j \braket{P_j|e^{-\beta H}|P_j}}.
\end{align}
To recover $\rho_\beta^{\rm (can)}$ accurately, one may sum over $2^N$ METTS with weight $p_i$, but performing such a calculation is usually intractable.
However, the METTS algorithm \cite{White09,SW10} enables one to generate a Markov chain of METTS and sample METTS according to the probability distribution $p_i$. Thus, there is no need to calculate $p_i$, and one can automatically sample only METTS with large $p_i$. Empirically, it is known that one only needs to sample roughly $10$ -- $100$ METTS to get accurate enough thermal expectation values of local observables. As a result, the most computationally costly part of this algorithm is to prepare each METTS by the imaginary time evolution of the product state.
As evaluated in \cite{White09,SW10}, the computation time to prepare a METTS with a matrix product state (MPS) ansatz \cite{schollwock05} scales as $N D^3 \beta$, where $D$ is the bond dimension of the MPS.

The METTS algorithm is indeed very efficient for simulating low temperature physics in 1D gapped systems
because it is expected that $D$ can be upper bounded by the bond dimension necessary to approximate the ground state \cite{White09,SW10}, which does not scale with $N$ or $\beta$ \cite{VC06,SWVC08}.

On the other hand, how well METTS behaves in 1D gapless systems is not clear at low temperature, while at high enough temperature it is expected to be similarly efficient to gapped systems.
There are many reasons for this. First, the arguments presented for gapped systems do not apply to gapless systems.
At the low temperature limit, METTS converge to the ground state, whose entanglement scales as $\CO(\log N)$ for a half of the whole 1D gapless system.
This implies that $D$ can be upper bounded by a polynomial of $N$, which does not manifest the efficiency of METTS when $N$ is large.
Therefore, one needs to understand how entanglement grows under imaginary time evolution for a generic METTS. \footnote{It is known that $D$ for {\it general} METTS is upper bounded by $\CO(e^{\sqrt{N\beta}})$ \cite{KAA20}, which does not manifest the efficiency of METTS at low temperatures. However, we only need to upper bound entanglement growth rate in {\it generic} METTS to evaluate the efficiency of the METTS algorithm. Therefore, there is a large room for improvement.}

Can we give a bound of entanglement growth under imaginary time evolution for generic METTS? Is METTS efficient for simulating low temperature physics of 1D gapless systems? To answer these questions, we focus on 1D gapless systems whose low energy sectors are well described by 2D conformal field theories. {After presenting some hints from CFT analyses,
we prove that for R\'enyi index $0<q\leq1$, the average entanglement R\'enyi entropy of a METTS decomposition is upper bounded by that of the thermofield double (TFD) state. Furthermore, we numerically show that the average entanglement R\'enyi entropy is about half of the analytic upper bound.
Based on these results, we
evaluate the bond dimension required to approximate METTS and the computation time for generating METTS.
In the end, we demonstrate the superiority of the METTS algorithm over the TFD algorithm \cite{FW05,BB15} (also called the ancilla method or purification method) at low temperatures in 1D gapless systems.
}

\textit{Hints from CFT computation.---}
While it is difficult to analytically compute R\'enyi entropies for all the METTS, we can do it for a special class of METTS with boundary conformal field theory (BCFT) computation. A BCFT is a CFT defined on a manifold with boundaries whose boundary conditions maximally preserve the conformal symmetries. Such boundary conditions, when regarded as quantum states, are called boundary states. One crucial feature of boundary states is that they are product states \cite{MRTW14}. Therefore, for a boundary state $\ket{B}$, $e^{-\beta H/2}\ket{B}$ turns out to be a METTS. 

For simplicity, let us consider a 1D system defined on an infinite line. The density matrix $e^{-\beta H/2}|B\rangle\langle B|e^{-\beta H/2}$ is realized by considering a path integral of an infinite strip with width $\beta$. The $q$-th entanglement R\'enyi entropy \footnote{For a density matrix $\rho_A$ defined on subsystem $A$, its $q$-th entanglement R\'enyi entropy is defined as $S^{(q)}_A = \frac{{\rm Tr}(\rho_A)^q}{1-q}$, and its entanglement entropy is defined as $S_A = -{\rm Tr}\left(\rho_A \log \rho_A\right)$. Note that $S_A = \lim_{q\rightarrow1} S_A^{(q)}$.} of a half line $A$ can be computed by evaluating the expectation value of a so-called twist operator inserted at the center of the strip \cite{CC04,HM13} and turns out to be 
\begin{align}\label{eq:HMEE}
    S_A^{(q)} = \frac{c}{12}\left(1+\frac{1}{q}\right) \log \left(\frac{2\beta}{\pi \epsilon}\right) + {\rm const.}, 
\end{align}
where $c$ is the CFT central charge, and $\epsilon$ is the UV cutoff. The constant term does not depend on $q$ but on the details of $\ket{B}$. Note that, except for this constant part, the above result is universal for any CFT and any $\ket{B}$. 
{Also note that $c$ roughly counts the degrees of freedom included in the CFT.} {For example, a free boson CFT has $c=1$ and an $n$-copy of free boson CFTs has $c=n$.}

From this result, one may expect that the average R\'enyi entropy of METTS has a similar scaling. With these as hints, let us move on to rigorous analyses. 

\textit{Analytic upper bound.---}
From now on, we will present a theorem which provides a rigorous upper bound for the average R\'enyi entropy of a METTS decomposition. 

We will utilize the TFD state.
Consider two identical copies $L$ and $R$ of the original system
($\mathcal{H} \cong \mathcal{H}_L \cong \mathcal{H}_R$).
Let $\mathcal{T}$ be an antiunitary operator that commutes with the Hamiltonian $H$.
Then the TFD state $\ket{\mathrm{TFD}} \in \mathcal{H}_L \otimes \mathcal{H}_R$ is defined as
\begin{align}
  \ket{\mathrm{TFD}}
  &= \sum_n e^{-\beta E_n/2} \ket{n}_L \otimes \ket{\overline{n}}_R,
\end{align}
where $\ket{n}$ are eigenstates of $H$ with energy $E_n$ and $\ket{\overline{\psi}}_R=\mathcal{T}\ket{\psi}_L$.
Since classical product states $\{\ket{P_i}\}_i$ form a complete orthonormal basis of $\mathcal{H}$,
we have
\begin{align}\label{eq:tfd_in_prod}
  \ket{\mathrm{TFD}}
  &= (e^{-\beta H/2} \otimes I_R) \sum_i \ket{P_i}_L \otimes \ket{\overline{P_i}}_R.
\end{align}

The TFD state is a (symmetric) purification of the canonical Gibbs state, i.e. 
  $\rho_{\beta}^\mathrm{(can)} \propto \mathrm{Tr}_R[\ket{\mathrm{TFD}}\bra{\mathrm{TFD}}].$
Therefore, one can also simulate finite temperature quantum systems with TFD states.
Dividing the original system
into $A$ and its complement $B$,
then each of the two copies $L$ and $R$
contains a copy of $A$.
We label them $A_L$ and $A_R$ respectively.

We find that the average entanglement R\'enyi entropy of $A$ over METTS is upper bounded by the entanglement R\'enyi entropy of $A_L A_R$ in $\ket{\rm TFD}$ associated with the same $\beta$, for R\'enyi index $0 < q \leq 1$.

\begin{theorem}
For $0<q\leq1$, there holds
\begin{align}\label{eq:upperbound_renyi}
  \overline{S_A^{(q)}}
  \equiv \sum_{i} p_i S_A^{(q)}(\ket{\mu_i})
  \leq S_{A_L A_R}^{(q)}(\ket{\mathrm{TFD}}),
\end{align}
{if $\ket{\overline{P_i}}_R$ are product states for $A_R$ and $B_R$.}
\end{theorem}

For generic quantum many-body systems, it is plausible to expect that $\ket{\overline{P_i}} \equiv \mathcal{T}\ket{P_i}$ is also a product state.
{For the models presented in this paper, $\mathcal{T}$ is simply the complex conjugate with respect to the spin basis, and this condition is satisfied.}

A detailed proof is presented in the Supple. Mat. \cite{Supple.}, which utilizes the concavity of $S^{(q)}$ for $0<q\leq 1$ since we used the concavity of $S^{(q)}$. This also means our proof is not valid for $q>1$. 
{However, we numerically find that the upper bound (\ref{eq:upperbound_renyi}) is valid even for $q>1$ in all the examples we study in this paper. It will be interesting to answer whether the result (\ref{eq:upperbound_renyi}) is valid for $q>1$ in general.} 

In fact, the $q=1$ case of this theorem follows directly from theorem 7 of \cite{WY22}, which was considered in a different context related to black hole physics. However, the extension to $0 < q < 1$ is necessary and sufficient to evaluate the required bond dimensions for an MPS to approximate METTS.

Let us then go back to 1D gapless systems defined on an infinite line and take the subsystem $A$ as a half of it. With CFT computations (See, e.g. \cite{Barthel17}), one finds 
\begin{align}\label{eq:rigorous_bound}
    \overline{S_{A}^{(q)}}
    \leq S_{A_L A_R}^{(q)}(\ket{\mathrm{TFD}}) = \frac{c}{6}\left(1+\frac{1}{q}\right) \log\left(\frac{\beta}{\pi \epsilon}\right),
\end{align}
for any CFT. This rigorous upper bound turns out to be twice of the right hand side of \eqref{eq:HMEE} at leading order. 

\begin{figure}
  \centering
  \includegraphics[width=7.8cm]{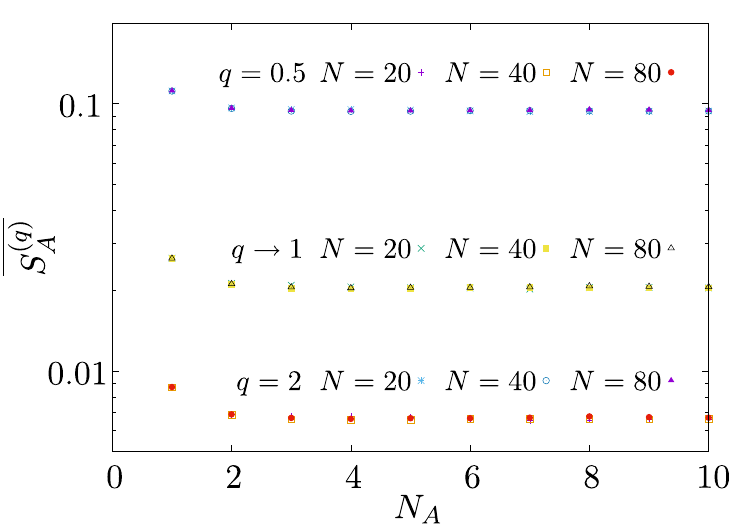}
  \caption{Average entanglement entropy of METTS
  between the left part (sites $n \leq N_A$) and the right part (sites $n > N_A$)
  as a function of the size $N_A$ of the subsystem $A$
  for the critical transverse-field Ising chain with $\nu=z$ and $\beta=4$. Imaginary-time evolution is carried out using the second-order Trotter decomposition with time step $\delta\tau=0.04$. The data is averaged over $10000$ samples.}
  \label{fig:TFI_EE-b_tau2qAxisZ}
\end{figure}

\begin{figure*}
  \centering
  \includegraphics[width=\linewidth]{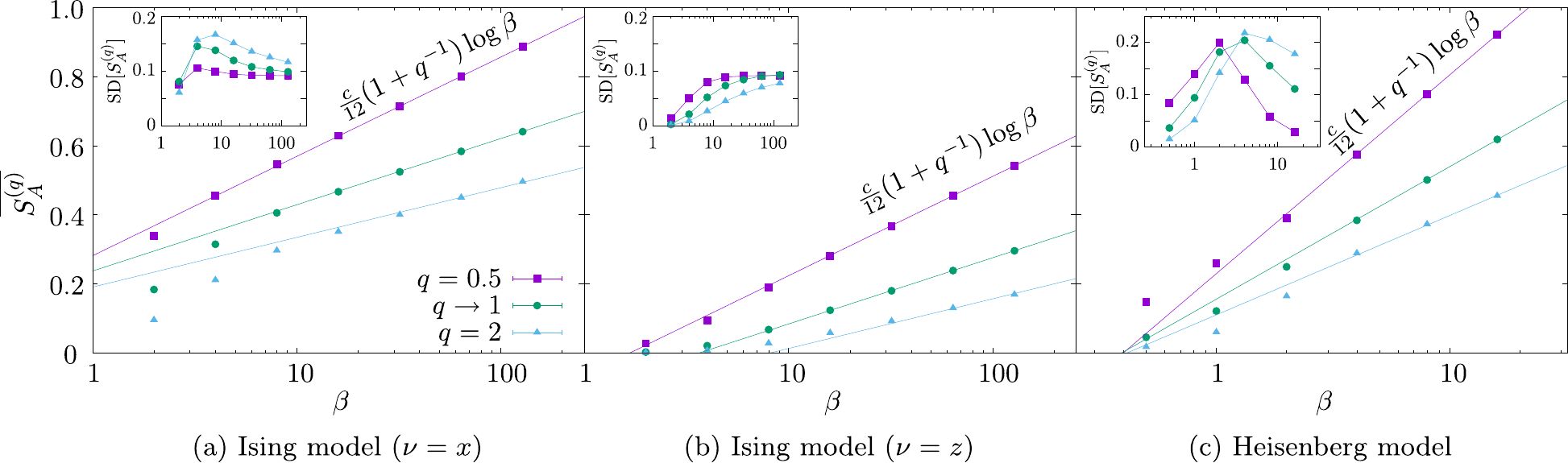}
  \caption{Growth of the average entanglement R\'enyi entropy of METTS for the half-chain $A=\{1,2,\cdots,N/2\}$
  in the critical transverse-field Ising model of length $N=640$ (a, b) and in the critical Heisenberg model with the next-nearest-neighbor interaction of length $N=1280$ (c). The insets show the standard deviation of the entanglement R\'enyi entropy. Imaginary-time evolution is carried out using the second-order Trotter decomposition with time step $\delta\tau=0.04$ for the Ising model and $\delta\tau=0.01$ for the Heisenberg model. The data is average over $10000$ samples for the Ising model and $5000$ samples for the Heisenberg model.}
  \label{fig:AverageEE}
\end{figure*}

\textit{Numerical study in spin systems.---}
After obtaining the rigorous bound \eqref{eq:rigorous_bound}, let us numerically study the average entanglement R\'enyi entropy of METTS in specific models. Our results show that it exhibits a universal behavior characterized only by the central charge of the associated CFT, similar to \eqref{eq:HMEE} for a regularized boundary state.

Numerical simulations are performed for two models with different central charges, symmetries and integrabilities. We take the spin quantization axis along $\nu$-direction ($\nu=x,y,z$) and choose eigenvectors of ${S}^\nu_{n}$ as $\{\ket{i_n}\}$ (which appears in the definition of $\ket{P_i}$). Here ${S}^{x,y,z}_{n}$ are the spin operators associated to the site $n$.

The first model is the transverse-field Ising chain (TFI) with the open boundary condition,
\begin{align}
  H = - \sum_{n=1}^{N-1} {S}^z_{n} {S}^z_{n+1}
  - \gamma \sum_{n=1}^{N} {S}^x_{n}, \label{eq:TFI}
\end{align}
which is integrable.
We set $\gamma=1/2$ so that the model is critical \cite{Sachdev11,CC04}. This theory is well described by CFT with central charge $c=1/2$. Since this Hamiltonian does not have $SU(2)$ spin rotation symmetry, entanglement properties of METTS depend on the choice of the quantization axis $\nu$. However, as will be seen later, the behavior of entanglement entropy at large $\beta$ is independent of $\nu$.

The second model is the spin-$1/2$ Heisenberg chain with the next-nearest-neighbor interaction,
\begin{align}
  H &= \sum_{n=1}^{N-1} ({S}^x_{n} {S}^x_{n+1} + {S}^y_{n} {S}^y_{n+1} + {S}^z_{n} {S}^z_{n+1}) \nonumber\\
  &+ J \sum_{n=1}^{N-2} ({S}^x_{n} {S}^x_{n+2} + {S}^y_{n} {S}^y_{n+2} + {S}^z_{n} {S}^z_{n+2}). \label{eq:NNNIHeis}
\end{align}
Again, we take the open boundary condition. We set $J=0.241167$ so that the model is critical \cite{ON92,NO94,Eggert96}. This theory is well described by CFT with central charge $c=1$ \cite{NO94}. This model is known to be not integrable \cite{HA93}.




As shown in Fig.~\ref{fig:TFI_EE-b_tau2qAxisZ}, the entanglement profiles is insensitive to the length of the whole spin chain $N$ or length of the interval subsystem. Therefore, it is sufficient to consider the case where the interval $A$ consists of the half-chain $A=\{1,2,\cdots,N/2\}$ for sufficiently large $N$ to investigate the $\beta$ dependence (see supplementary material \cite{Supple.} and also references \cite{WY22,Rastegin2011} therein). Then, in Fig.~\ref{fig:AverageEE}, we show the average entanglement R\'enyi entropy for the half-chain as a function of $\beta$. For $\beta \gg 1$, we find
\begin{align} \label{eq:EE_METTS}
  \overline{S_A^{(q)}}
  \simeq \frac{c}{12} \left(1+\frac{1}{q}\right) \log \beta + S_0^{(q)},
\end{align}
where $S_0^{(q)}$ is a non-universal constant.
Furthermore, the standard deviation of the entanglement entropy of METTS is bounded by a constant and is negligible in the low temperature limit as compared to the average. That is, at low temperatures, the entanglement R\'enyi entropy of METTS coincides with the entropy of the regularized boundary state \eqref{eq:HMEE} and a half of our upper bound by the entropy of the TFD state \eqref{eq:rigorous_bound} at the leading order of $\beta$. This result is consistent with the intuition obtained from the discussion in the low temperature limit pointed out in Ref.~\cite{BB15}. However, it is not immediately evident that the entanglement of METTS is half of the TFD state even in the finite temperature region since the entanglement of the imaginary-time evolved state significantly depends on the choice of the initial state.

\textit{Computation time in the METTS algorithm.---}
The area law and the logarithmically slow growth of the entanglement R\'enyi entropy \eqref{eq:EE_METTS} for $0<q<1$ implies that METTS can be efficiently represented by an MPS ansatz \cite{VC06,SWVC08}. Hence it is expected that the METTS algorithm can be used to reach considerably low temperatures even for critical spin chains. With the results obtained above, let us evaluate the computation time in the METTS algorithm for simulating the canonical Gibbs state.

Consider a Schmidt decomposition of the METTS $\ket{\mu_i} = \sum_k \lambda_k \ket{k}_{A} \ket{k}_{B}$ with respect to the bipartition $A=\{1,2,\cdots,N_A\}$ and its complement $B$. Suppose that the Schmidt coefficients $\lambda_i$'s are sorted in decreasing order, i.e., $\lambda_1 \geq \lambda_2 \geq \cdots$. The truncation error for an MPS approximation of bond dimension $D$ is defined as $\varepsilon \equiv \sum_{k>D} {\lambda_k}^2$. As shown in \cite{VC06}, $\varepsilon$ measures an error in an MPS approximation in the following sense: there exists an MPS $\ket{\psi_D}$ of bond dimension $D$ such that ${\|\ket{\mu_i}-\ket{\psi_D}\|_2}^2 \leq 2 \sum_{N_A=1}^{N-1} \varepsilon$. Furthermore, $\varepsilon$ can be bounded from above by
\begin{align}
    \log \varepsilon \leq \frac{1-q}{q} \left( S_A^{(q)} - \log \frac{D}{1-q} \right),
\end{align}
for any $0<q<1$ \cite{VC06}. By inverting this inequality, we find an upper bound on the bond dimension necessary to approximate the METTS within the truncation error $\varepsilon$:
\begin{align}
  \log D \leq S_A^{(q)}-\frac{q}{1-q}\log \varepsilon + \log (1-q).
\end{align}
Substituting the numerical result \eqref{eq:EE_METTS}, we get an upper bound on $D$ for generic METTS scales as
\begin{align} \label{eq:D-bound_METTS}
  \CO(N^0 \beta^{\frac{c}{12}(1+q^{-1})}),
\end{align}
for fixed $\varepsilon$ and $q$. Note that the constant factor depends on $\varepsilon$ and $q$, and hence it does not necessarily mean that the highest efficiency is achieved at $q=1$. On the other hand, also for the TFD state, there is an upper bound on $D$ scales as \cite{Barthel17}
\begin{align} \label{eq:D-bound_TDF}
  \CO(N^0 \beta^{\frac{c}{6}(1+q^{-1})}).
\end{align}

\begin{figure}
  \centering
  \includegraphics[width=7.8cm]{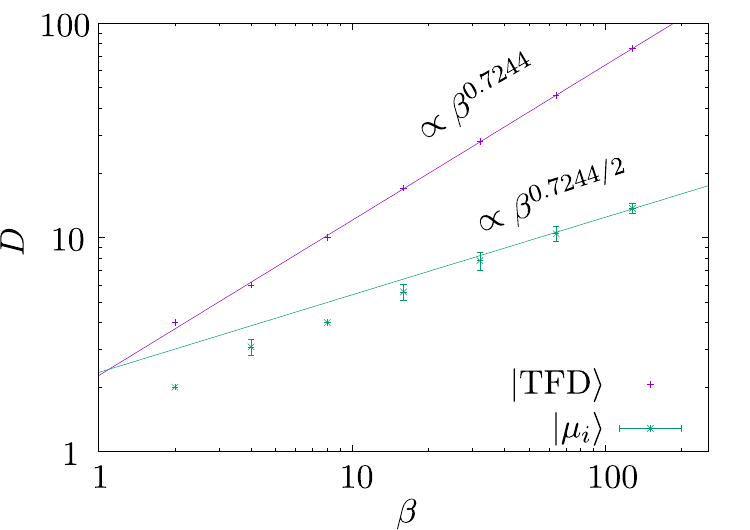}
  \caption{Minimum bond dimension $D$ necessary to approximate the TFD state and METTS with the truncation error less than $10^{-10}$ for the critical transverse-field Ising chain with $N=640$ sites. The data points for METTS are the average of $D$ of $10000$ samples. The error bars indicate the standard deviation. We set the quantization axis $\nu$ to $z$. Imaginary-time evolution is carried out using the second-order Trotter decomposition with time step $\delta\tau=0.04$.}
  \label{fig:TFI_beta-bondDim_L640}
\end{figure}

Then we numerically check the bond dimension necessary to approximate the TFD state and METTS. In Fig.~\ref{fig:TFI_beta-bondDim_L640}, we show the minimum bond dimension needed to achieve the truncation error $\varepsilon$ less than $10^{-10}$ for the TFD state and METTS as a function of $\beta$. We can confirm that the exponent on $\beta$ for the TFD state is just twice of that for generic METTS in the low temperature limit. This is consistent with Eqs.~\eqref{eq:D-bound_METTS} and \eqref{eq:D-bound_TDF}.

According to \cite{SW10}, the computation time of the imaginary time evolution for $\beta$ scales as $ND^3\beta$. Therefore, we can efficiently produce METTS at sufficiently low temperatures with the computation time that scales as a polynomial in $\beta$ with an exponent smaller than that of the TFD state.

\begin{figure}
  \centering
  \includegraphics[width=8cm]{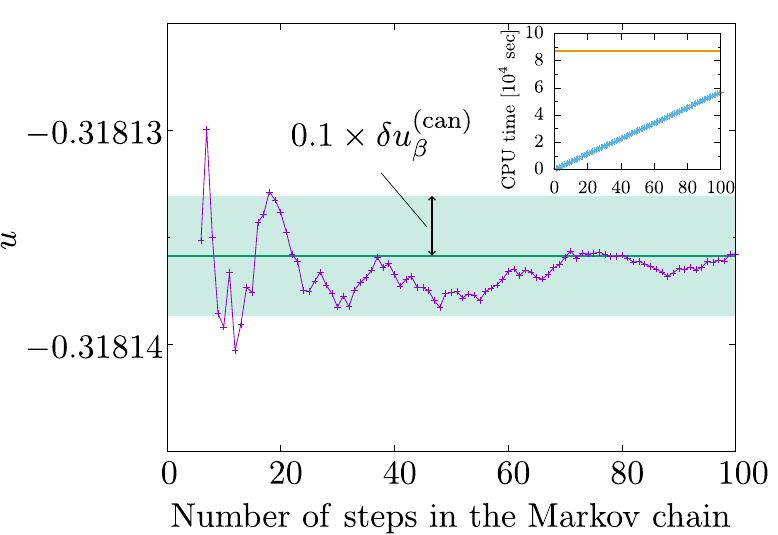}
  \caption{Typical trajectory of the cumulative moving average of the energy per site $u=E/N$ in the Markov chain of the METTS simulation for the critical transverse-field Ising chain with $N=640$ sites at $\beta=128$. We set the quantization axis $\nu$ to $z$ for the odd-numbered steps and $x$ for the even-numbered steps to reduce correlations between successive samples \cite{BB17}. To avoid initial transients, we discard the first $5$ samples when calculating the sample averages. The solid line shows the result of the TFD approach, considered exact, and the shaded region shows a range of $0.1$ times the standard deviation ${\delta u}_\beta^{\rm (can)}$ of $u$ in $\rho_\beta^{\rm(can)}$ from the exact value. The inset shows the total computation time in the METTS simulation as a function of the number of steps in the Markov chain. The solid line shows the computation time in the TFD simulation.}
  \label{fig:TFI_CMA-CPUtime_L640}
\end{figure}

We finally demonstrate the superiority of the METTS algorithm over the TFD algorithm at low temperatures by comparing the computation time in actual numerical simulations. To calculate expectation values in the canonical Gibbs state $\rho_\beta^{\rm(can)}$ via the METTS algorithm, we generate a Markov chain of METTS so that the long-time average coincides with $\rho_\beta^{\rm(can)}$ and then calculate the average of the expectation values in METTS sampled from that chain.
Fig.~\ref{fig:TFI_CMA-CPUtime_L640} depicts a typical trajectory of the cumulative moving average of the energy per site $u=E/N$ in METTS and the computation time in the METTS simulation. It can be seen that it only takes about $20$ steps for sample averages to converge within a range of $0.1$ times the standard deviation ${\delta u}_\beta^{\rm (can)}$ of $u$ in $\rho_\beta^{\rm(can)}$ from the exact value. In other words, with only a few samples, we can correctly obtain the energy with an accuracy sufficiently smaller than the uncertainty inherent in the ensemble. Furthermore, we also observe that the total computation time for reaching that step is only about $1/8$ of that for the TFD algorithm. We therefore conclude that the METTS algorithm can provide a speedup
over the TFD algorithm when analyzing low temperature states in 1D gapless systems.
However, in both algorithms, it is possible to reduce the computation time by increasing the Trotter error or truncation error within an allowable error range, and in such case the TFD algorithm outperform the METTS algorithm~\cite{BB15}. Nevertheless, these errors are difficult to control, compared to statistical errors, which can be easily controlled by simply increasing the number of samples. In addition, in the case of the METTS algorithm, it is possible to parallelize the sampling to reduce the computation time.

\textit{Conclusions.---}
In this paper, we focus on evaluating the computational efficiency of the METTS algorithm for simulating low temperature thermal equilibrium states in 1D gapless systems. We find it is efficient in the sense that the computation time scales as a polynomial function of $\beta$ and has a significant superiority compared to the TFD method. These are based on both the analytic upper bound and numerical investigations on the average entanglement R\'enyi entropy of METTS, which are motivated by a hint obtained in a BCFT computation.

We would like to emphasize that, although it was well-known in the experience that a METTS usually requires fewer bond dimensions than a TFD state and the METTS algorithm takes less time than the TFD method in some cases, why and to what extent this is true was not well-understood so far, both for gapped systems and gapless systems. More than justifying this experimental rule, our results also demonstrated that the speedup of METTS compared to TFD for simulating 1D gapless systems at low temperatures is parametrically large, i.e., the speedup gets larger as $\beta$ gets larger, different from 1D gapped systems.


From a theoretical point of view, the average entanglement entropy of METTS is also capable of upper bounding the entanglement of formation for the canonical Gibbs state, which is defined as the minimum average entanglement over all of its possible pure state decompositions.
This further bounds operational entanglement measures, such as distillable entanglement and entanglement cost. It is expected that the METTS decomposition is quite close to being optimal. Thus, it is possible to obtain a tight bound on the entanglement of formation with the average entanglement of METTS. Therefore, by utilizing the results of this work, one can investigate the amount of entanglement in canonical Gibbs states.

Last but not least, METTS are low-entanglement atypical states mimicking thermal equilibrium. Translated to the language of quantum gravity via the AdS/CFT correspondence, this means METTS may be understood as black hole microstates which have nontrivial structures near the horizon \cite{WY22}. Therefore, the statistical features of METTS may play an important role in understanding the microstate physics of black holes.

{\bf Acknowledgements}
We are grateful to Tomotaka Kuwahara, Hiroyasu Tajima, Tadashi Takayanagi and Yantao Wu for useful discussions. 
We would like to especially thank Yoshifumi Nakata and Yantao Wu for careful reading and valuable comments on a draft of this paper.
YK is supported by the Brinson Prize Fellowship at Caltech and the U.S. Department of Energy, Office of Science, Office of High Energy Physics, under Award Number DE-SC0011632. KT is supported by Grant-in-Aid for Early-Career Scientists No.~21K13920 and Grant-in-Aid for Transformative Research Areas (A) No.~22H05265. ZW is supported by the Society of Fellows at Harvard University.

\appendix

\bibliography{METTS}

\begin{thebibliography}{23}%
\makeatletter
\providecommand \@ifxundefined [1]{%
 \@ifx{#1\undefined}
}%
\providecommand \@ifnum [1]{%
 \ifnum #1\expandafter \@firstoftwo
 \else \expandafter \@secondoftwo
 \fi
}%
\providecommand \@ifx [1]{%
 \ifx #1\expandafter \@firstoftwo
 \else \expandafter \@secondoftwo
 \fi
}%
\providecommand \natexlab [1]{#1}%
\providecommand \enquote  [1]{``#1''}%
\providecommand \bibnamefont  [1]{#1}%
\providecommand \bibfnamefont [1]{#1}%
\providecommand \citenamefont [1]{#1}%
\providecommand \href@noop [0]{\@secondoftwo}%
\providecommand \href [0]{\begingroup \@sanitize@url \@href}%
\providecommand \@href[1]{\@@startlink{#1}\@@href}%
\providecommand \@@href[1]{\endgroup#1\@@endlink}%
\providecommand \@sanitize@url [0]{\catcode `\\12\catcode `\$12\catcode `\&12\catcode `\#12\catcode `\^12\catcode `\_12\catcode `\%12\relax}%
\providecommand \@@startlink[1]{}%
\providecommand \@@endlink[0]{}%
\providecommand \url  [0]{\begingroup\@sanitize@url \@url }%
\providecommand \@url [1]{\endgroup\@href {#1}{\urlprefix }}%
\providecommand \urlprefix  [0]{URL }%
\providecommand \Eprint [0]{\href }%
\providecommand \doibase [0]{https://doi.org/}%
\providecommand \selectlanguage [0]{\@gobble}%
\providecommand \bibinfo  [0]{\@secondoftwo}%
\providecommand \bibfield  [0]{\@secondoftwo}%
\providecommand \translation [1]{[#1]}%
\providecommand \BibitemOpen [0]{}%
\providecommand \bibitemStop [0]{}%
\providecommand \bibitemNoStop [0]{.\EOS\space}%
\providecommand \EOS [0]{\spacefactor3000\relax}%
\providecommand \BibitemShut  [1]{\csname bibitem#1\endcsname}%
\let\auto@bib@innerbib\@empty
\bibitem [{\citenamefont {White}(2009)}]{White09}%
  \BibitemOpen
  \bibfield  {author} {\bibinfo {author} {\bibfnamefont {S.~R.}\ \bibnamefont {White}},\ }\bibfield  {title} {\bibinfo {title} {Minimally entangled typical quantum states at finite temperature},\ }\bibfield  {journal} {\bibinfo  {journal} {Physical Review Letters}\ }\textbf {\bibinfo {volume} {102}},\ \href {https://doi.org/10.1103/physrevlett.102.190601} {10.1103/physrevlett.102.190601} (\bibinfo {year} {2009}),\ \Eprint {https://arxiv.org/abs/0902.4475} {arXiv:0902.4475 [cond-mat]} \BibitemShut {NoStop}%
\bibitem [{\citenamefont {Stoudenmire}\ and\ \citenamefont {White}(2010)}]{SW10}%
  \BibitemOpen
  \bibfield  {author} {\bibinfo {author} {\bibfnamefont {E.~M.}\ \bibnamefont {Stoudenmire}}\ and\ \bibinfo {author} {\bibfnamefont {S.~R.}\ \bibnamefont {White}},\ }\bibfield  {title} {\bibinfo {title} {Minimally entangled typical thermal state algorithms},\ }\href {https://doi.org/10.1088/1367-2630/12/5/055026} {\bibfield  {journal} {\bibinfo  {journal} {New Journal of Physics}\ }\textbf {\bibinfo {volume} {12}},\ \bibinfo {pages} {055026} (\bibinfo {year} {2010})},\ \Eprint {https://arxiv.org/abs/1002.1305} {arXiv:1002.1305 [cond-mat]} \BibitemShut {NoStop}%
\bibitem [{\citenamefont {Schollwöck}(2005)}]{schollwock05}%
  \BibitemOpen
  \bibfield  {author} {\bibinfo {author} {\bibfnamefont {U.}~\bibnamefont {Schollwöck}},\ }\bibfield  {title} {\bibinfo {title} {The density-matrix renormalization group},\ }\href {https://doi.org/10.1103/RevModPhys.77.259} {\bibfield  {journal} {\bibinfo  {journal} {Rev. Mod. Phys.}\ }\textbf {\bibinfo {volume} {77}},\ \bibinfo {pages} {259} (\bibinfo {year} {2005})}\BibitemShut {NoStop}%
\bibitem [{\citenamefont {Verstraete}\ and\ \citenamefont {Cirac}(2006)}]{VC06}%
  \BibitemOpen
  \bibfield  {author} {\bibinfo {author} {\bibfnamefont {F.}~\bibnamefont {Verstraete}}\ and\ \bibinfo {author} {\bibfnamefont {J.~I.}\ \bibnamefont {Cirac}},\ }\bibfield  {title} {\bibinfo {title} {{Matrix product states represent ground states faithfully}},\ }\href {https://doi.org/10.1103/PhysRevB.73.094423} {\bibfield  {journal} {\bibinfo  {journal} {Phys. Rev. B - Condens. Matter Mater. Phys.}\ }\textbf {\bibinfo {volume} {73}},\ \bibinfo {pages} {094423} (\bibinfo {year} {2006})},\ \Eprint {https://arxiv.org/abs/0505140} {arXiv:0505140 [cond-mat]} \BibitemShut {NoStop}%
\bibitem [{\citenamefont {Schuch}\ \emph {et~al.}(2008)\citenamefont {Schuch}, \citenamefont {Wolf}, \citenamefont {Verstraete},\ and\ \citenamefont {Cirac}}]{SWVC08}%
  \BibitemOpen
  \bibfield  {author} {\bibinfo {author} {\bibfnamefont {N.}~\bibnamefont {Schuch}}, \bibinfo {author} {\bibfnamefont {M.~M.}\ \bibnamefont {Wolf}}, \bibinfo {author} {\bibfnamefont {F.}~\bibnamefont {Verstraete}},\ and\ \bibinfo {author} {\bibfnamefont {J.~I.}\ \bibnamefont {Cirac}},\ }\bibfield  {title} {\bibinfo {title} {{Entropy scaling and simulability by matrix product states}},\ }\href {https://doi.org/10.1103/PhysRevLett.100.030504} {\bibfield  {journal} {\bibinfo  {journal} {Phys. Rev. Lett.}\ }\textbf {\bibinfo {volume} {100}},\ \bibinfo {pages} {030504} (\bibinfo {year} {2008})},\ \Eprint {https://arxiv.org/abs/0705.0292} {arXiv:0705.0292} \BibitemShut {NoStop}%
\bibitem [{Note1()}]{Note1}%
  \BibitemOpen
  \bibinfo {note} {It is known that $D$ for {\protect \it general} METTS is upper bounded by ${\protect \mathcal {O}}(e^{\protect \sqrt {N\beta }})$ \cite {KAA20}, which does not manifest the efficiency of METTS at low temperatures. However, we only need to upper bound entanglement growth rate in {\protect \it generic} METTS to evaluate the efficiency of the METTS algorithm. Therefore, there is a large room for improvement.}\BibitemShut {Stop}%
\bibitem [{\citenamefont {Feiguin}\ and\ \citenamefont {White}(2005)}]{FW05}%
  \BibitemOpen
  \bibfield  {author} {\bibinfo {author} {\bibfnamefont {A.~E.}\ \bibnamefont {Feiguin}}\ and\ \bibinfo {author} {\bibfnamefont {S.~R.}\ \bibnamefont {White}},\ }\bibfield  {title} {\bibinfo {title} {{Finite-temperature density matrix renormalization using an enlarged Hilbert space}},\ }\href {https://doi.org/10.1103/PhysRevB.72.220401} {\bibfield  {journal} {\bibinfo  {journal} {Phys. Rev. B}\ }\textbf {\bibinfo {volume} {72}},\ \bibinfo {pages} {220401} (\bibinfo {year} {2005})}\BibitemShut {NoStop}%
\bibitem [{\citenamefont {Binder}\ and\ \citenamefont {Barthel}(2015)}]{BB15}%
  \BibitemOpen
  \bibfield  {author} {\bibinfo {author} {\bibfnamefont {M.}~\bibnamefont {Binder}}\ and\ \bibinfo {author} {\bibfnamefont {T.}~\bibnamefont {Barthel}},\ }\bibfield  {title} {\bibinfo {title} {Minimally entangled typical thermal states versus matrix product purifications for the simulation of equilibrium states and time evolution},\ }\href {https://doi.org/10.1103/PhysRevB.92.125119} {\bibfield  {journal} {\bibinfo  {journal} {Physical Review B}\ }\textbf {\bibinfo {volume} {92}},\ \bibinfo {pages} {125119} (\bibinfo {year} {2015})},\ \bibinfo {note} {american Physical Society}\BibitemShut {NoStop}%
\bibitem [{\citenamefont {Miyaji}\ \emph {et~al.}(2015)\citenamefont {Miyaji}, \citenamefont {Ryu}, \citenamefont {Takayanagi},\ and\ \citenamefont {Wen}}]{MRTW14}%
  \BibitemOpen
  \bibfield  {author} {\bibinfo {author} {\bibfnamefont {M.}~\bibnamefont {Miyaji}}, \bibinfo {author} {\bibfnamefont {S.}~\bibnamefont {Ryu}}, \bibinfo {author} {\bibfnamefont {T.}~\bibnamefont {Takayanagi}},\ and\ \bibinfo {author} {\bibfnamefont {X.}~\bibnamefont {Wen}},\ }\bibfield  {title} {\bibinfo {title} {{Boundary States as Holographic Duals of Trivial Spacetimes}},\ }\href {https://doi.org/10.1007/JHEP05(2015)152} {\bibfield  {journal} {\bibinfo  {journal} {JHEP}\ }\textbf {\bibinfo {volume} {05}},\ \bibinfo {pages} {152}},\ \Eprint {https://arxiv.org/abs/1412.6226} {arXiv:1412.6226 [hep-th]} \BibitemShut {NoStop}%
\bibitem [{Note2()}]{Note2}%
  \BibitemOpen
  \bibinfo {note} {For a density matrix $\rho _A$ defined on subsystem $A$, its $q$-th entanglement R\'enyi entropy is defined as $S^{(q)}_A = \protect \frac {{\protect \rm Tr}(\rho _A)^q}{1-q}$, and its entanglement entropy is defined as $S_A = -{\protect \rm Tr}\left (\rho _A \log \rho _A\right )$. Note that $S_A = \lim _{q\rightarrow 1} S_A^{(q)}$.}\BibitemShut {Stop}%
\bibitem [{\citenamefont {Calabrese}\ and\ \citenamefont {Cardy}(2004)}]{CC04}%
  \BibitemOpen
  \bibfield  {author} {\bibinfo {author} {\bibfnamefont {P.}~\bibnamefont {Calabrese}}\ and\ \bibinfo {author} {\bibfnamefont {J.~L.}\ \bibnamefont {Cardy}},\ }\bibfield  {title} {\bibinfo {title} {{Entanglement entropy and quantum field theory}},\ }\href {https://doi.org/10.1088/1742-5468/2004/06/P06002} {\bibfield  {journal} {\bibinfo  {journal} {J. Stat. Mech.}\ }\textbf {\bibinfo {volume} {0406}},\ \bibinfo {pages} {P06002} (\bibinfo {year} {2004})},\ \Eprint {https://arxiv.org/abs/hep-th/0405152} {arXiv:hep-th/0405152} \BibitemShut {NoStop}%
\bibitem [{\citenamefont {Hartman}\ and\ \citenamefont {Maldacena}(2013)}]{HM13}%
  \BibitemOpen
  \bibfield  {author} {\bibinfo {author} {\bibfnamefont {T.}~\bibnamefont {Hartman}}\ and\ \bibinfo {author} {\bibfnamefont {J.}~\bibnamefont {Maldacena}},\ }\bibfield  {title} {\bibinfo {title} {{Time Evolution of Entanglement Entropy from Black Hole Interiors}},\ }\href {https://doi.org/10.1007/JHEP05(2013)014} {\bibfield  {journal} {\bibinfo  {journal} {JHEP}\ }\textbf {\bibinfo {volume} {05}},\ \bibinfo {pages} {014}},\ \Eprint {https://arxiv.org/abs/1303.1080} {arXiv:1303.1080 [hep-th]} \BibitemShut {NoStop}%
\bibitem [{Sup()}]{Supple.}%
  \BibitemOpen
  \bibinfo {note} {Supplementary materials.}\BibitemShut {Stop}%
\bibitem [{\citenamefont {Wei}\ and\ \citenamefont {Yoneta}(2024)}]{WY22}%
  \BibitemOpen
  \bibfield  {author} {\bibinfo {author} {\bibfnamefont {Z.}~\bibnamefont {Wei}}\ and\ \bibinfo {author} {\bibfnamefont {Y.}~\bibnamefont {Yoneta}},\ }\bibfield  {title} {\bibinfo {title} {{Counting atypical black hole microstates from entanglement wedges}},\ }\href {https://doi.org/10.1007/JHEP05(2024)251} {\bibfield  {journal} {\bibinfo  {journal} {JHEP}\ }\textbf {\bibinfo {volume} {05}},\ \bibinfo {pages} {251}},\ \Eprint {https://arxiv.org/abs/2211.11787} {arXiv:2211.11787 [hep-th]} \BibitemShut {NoStop}%
\bibitem [{\citenamefont {Barthel}(2017)}]{Barthel17}%
  \BibitemOpen
  \bibfield  {author} {\bibinfo {author} {\bibfnamefont {T.}~\bibnamefont {Barthel}},\ }\bibfield  {title} {\bibinfo {title} {One-dimensional quantum systems at finite temperatures can be simulated efficiently on classical computers},\ }\href@noop {} {\  (\bibinfo {year} {2017})},\ \Eprint {https://arxiv.org/abs/1708.09349} {arXiv:1708.09349 [quant-ph]} \BibitemShut {NoStop}%
\bibitem [{\citenamefont {Sachdev}(2011)}]{Sachdev11}%
  \BibitemOpen
  \bibfield  {author} {\bibinfo {author} {\bibfnamefont {S.}~\bibnamefont {Sachdev}},\ }\href@noop {} {\emph {\bibinfo {title} {{Quantum Phase Transitions}}}}\ (\bibinfo  {publisher} {Cambridge University Press, Cambridge, England},\ \bibinfo {year} {2011})\BibitemShut {NoStop}%
\bibitem [{\citenamefont {Okamoto}\ and\ \citenamefont {Nomura}(1992)}]{ON92}%
  \BibitemOpen
  \bibfield  {author} {\bibinfo {author} {\bibfnamefont {K.}~\bibnamefont {Okamoto}}\ and\ \bibinfo {author} {\bibfnamefont {K.}~\bibnamefont {Nomura}},\ }\bibfield  {title} {\bibinfo {title} {{Fluid-dimer critical point in S = 1/2 antiferromagnetic Heisenberg chain with next nearest neighbor interactions}},\ }\href {https://doi.org/10.1016/0375-9601(92)90823-5} {\bibfield  {journal} {\bibinfo  {journal} {Phys. Lett. A}\ }\textbf {\bibinfo {volume} {169}},\ \bibinfo {pages} {433} (\bibinfo {year} {1992})}\BibitemShut {NoStop}%
\bibitem [{\citenamefont {Nomura}\ and\ \citenamefont {Okamoto}(1994)}]{NO94}%
  \BibitemOpen
  \bibfield  {author} {\bibinfo {author} {\bibfnamefont {K.}~\bibnamefont {Nomura}}\ and\ \bibinfo {author} {\bibfnamefont {K.}~\bibnamefont {Okamoto}},\ }\bibfield  {title} {\bibinfo {title} {{Critical properties of S= 1/2 antiferromagnetic XXZ chain with next-nearest-neighbour interactions}},\ }\href {https://doi.org/10.1088/0305-4470/27/17/012} {\bibfield  {journal} {\bibinfo  {journal} {J. Phys. A. Math. Gen.}\ }\textbf {\bibinfo {volume} {27}},\ \bibinfo {pages} {5773} (\bibinfo {year} {1994})}\BibitemShut {NoStop}%
\bibitem [{\citenamefont {Eggert}(1996)}]{Eggert96}%
  \BibitemOpen
  \bibfield  {author} {\bibinfo {author} {\bibfnamefont {S.}~\bibnamefont {Eggert}},\ }\bibfield  {title} {\bibinfo {title} {{Numerical evidence for multiplicative logarithmic corrections from marginal operators}},\ }\href {https://doi.org/10.1103/PhysRevB.54.R9612} {\bibfield  {journal} {\bibinfo  {journal} {Phys. Rev. B - Condens. Matter Mater. Phys.}\ }\textbf {\bibinfo {volume} {54}},\ \bibinfo {pages} {R9612} (\bibinfo {year} {1996})},\ \Eprint {https://arxiv.org/abs/9602026} {arXiv:9602026 [cond-mat]} \BibitemShut {NoStop}%
\bibitem [{\citenamefont {Hsu}\ and\ \citenamefont {{Angle's D'Auriac}}(1993)}]{HA93}%
  \BibitemOpen
  \bibfield  {author} {\bibinfo {author} {\bibfnamefont {T.~C.}\ \bibnamefont {Hsu}}\ and\ \bibinfo {author} {\bibfnamefont {J.~C.}\ \bibnamefont {{Angle's D'Auriac}}},\ }\bibfield  {title} {\bibinfo {title} {{Level repulsion in integrable and almost-integrable quantum spin models}},\ }\href {https://doi.org/10.1103/PhysRevB.47.14291} {\bibfield  {journal} {\bibinfo  {journal} {Phys. Rev. B}\ }\textbf {\bibinfo {volume} {47}},\ \bibinfo {pages} {14291} (\bibinfo {year} {1993})}\BibitemShut {NoStop}%
\bibitem [{\citenamefont {Rastegin}(2011)}]{Rastegin2011}%
  \BibitemOpen
  \bibfield  {author} {\bibinfo {author} {\bibfnamefont {A.~E.}\ \bibnamefont {Rastegin}},\ }\bibfield  {title} {\bibinfo {title} {{Some General Properties of Unified Entropies}},\ }\href {https://doi.org/10.1007/s10955-011-0231-x} {\bibfield  {journal} {\bibinfo  {journal} {J. Stat. Phys.}\ }\textbf {\bibinfo {volume} {143}},\ \bibinfo {pages} {1120} (\bibinfo {year} {2011})},\ \Eprint {https://arxiv.org/abs/1012.5356} {arXiv:1012.5356} \BibitemShut {NoStop}%
\bibitem [{\citenamefont {Binder}\ and\ \citenamefont {Barthel}(2017)}]{BB17}%
  \BibitemOpen
  \bibfield  {author} {\bibinfo {author} {\bibfnamefont {M.}~\bibnamefont {Binder}}\ and\ \bibinfo {author} {\bibfnamefont {T.}~\bibnamefont {Barthel}},\ }\bibfield  {title} {\bibinfo {title} {Symmetric minimally entangled typical thermal states for canonical and grand-canonical ensembles},\ }\href {https://doi.org/10.1103/PhysRevB.95.195148} {\bibfield  {journal} {\bibinfo  {journal} {Physical Review B}\ }\textbf {\bibinfo {volume} {95}},\ \bibinfo {pages} {195148} (\bibinfo {year} {2017})},\ \bibinfo {note} {american Physical Society}\BibitemShut {NoStop}%
\bibitem [{\citenamefont {Kuwahara}\ \emph {et~al.}(2021)\citenamefont {Kuwahara}, \citenamefont {Alhambra},\ and\ \citenamefont {Anshu}}]{KAA20}%
  \BibitemOpen
  \bibfield  {author} {\bibinfo {author} {\bibfnamefont {T.}~\bibnamefont {Kuwahara}}, \bibinfo {author} {\bibfnamefont {{\'{A}}.~M.}\ \bibnamefont {Alhambra}},\ and\ \bibinfo {author} {\bibfnamefont {A.}~\bibnamefont {Anshu}},\ }\bibfield  {title} {\bibinfo {title} {Improved thermal area law and quasilinear time algorithm for quantum gibbs states},\ }\bibfield  {journal} {\bibinfo  {journal} {Physical Review X}\ }\textbf {\bibinfo {volume} {11}},\ \href {https://doi.org/10.1103/physrevx.11.011047} {10.1103/physrevx.11.011047} (\bibinfo {year} {2021}),\ \Eprint {https://arxiv.org/abs/2007.11174} {arXiv:2007.11174 [quant-ph]} \BibitemShut {NoStop}%
\end{thebibliography}%

\newcommand{\beginsupplement}{%
	\setcounter{table}{0}
	\renewcommand{\thetable}{S\arabic{table}}%
	\setcounter{figure}{0}
	\renewcommand{\thefigure}{S\arabic{figure}}%
	\setcounter{section}{0}
	\renewcommand{\thesection}{\Roman{section}}%
	\setcounter{equation}{0}
	\renewcommand{\theequation}{S\arabic{equation}}%
}
\clearpage

\onecolumngrid

\beginsupplement

\begin{center}
	\textbf{\large Supplementary Materials for `` Efficient Simulation of Low Temperature Physics in One-Dimensional Gapless Systems  ''}
\end{center}
\vspace{2mm}

\section{Proof of Theorem~1}
Here we present a detailed proof for Theorem 1. appearing in the main text. As explained in the main text, the proof utilizes concavity of R\'enyi entropy for R\'enyi index $0<q<1$ and that of von Neumann entropy. The proof presented here is parallel to the proof of theorem 7 in \cite{WY22}, which is a related statement for von Neumann entropy. 

\begin{proof}
Take sets of product states
$\{\ket{P_i}_A\}_i$ and $\{\ket{P_j}_B\}_j$
of subsystems $A$ and $B$, respectively,
so that $\{\ket{P_i}\}_i = \{\ket{P_i}_A \otimes \ket{P_j}_B\}_{i,j}$.
Consider the projection
\begin{align}
  K_j &= I_{A_L} \otimes I_{B_L} \otimes I_{A_R} \otimes \ket{\overline{P_j}}\bra{\overline{P_j}}_{B_R}.
\end{align}
Then, from the completeness of $K_j$, we get
\begin{align}
  &\mathrm{Tr}_{B_L B_R} \left[ \ket{\mathrm{TFD}}\bra{\mathrm{TFD}} \right] \nonumber\\
  =& \sum_j \mathrm{Tr}_{B_L B_R} \left[ K_j \ket{\mathrm{TFD}}\bra{\mathrm{TFD}} {K_j}^\dagger \right].
\end{align}
Therefore, we have
\begin{align}
  S_{A_L A_R}^{(q)}(\ket{\mathrm{TFD}})
  &= S^{(q)} \left( \sum_j s_j \rho_{A_L A_R}^{K_j\ket{\mathrm{TFD}}} \right),
\end{align}
where
\begin{align}
  s_j = \frac{\braket{\mathrm{TFD}|K_j|\mathrm{TFD}}}{\braket{\mathrm{TFD}|\mathrm{TFD}}}.
\end{align}
Further, using the matrix concavity of the R\'enyi entropy $S^{(q)}$ with $0<q<1$ \cite{Rastegin2011} and with $q=1$ (i.e., the concavity of the von Neumann entropy), we obtain
\begin{align}\label{eq:concavity}
  S_{A_L A_R}^{(q)}(\ket{\mathrm{TFD}})
  &\geq  \sum_j s_j S_{A_L A_R}^{(q)}(K_j\ket{\mathrm{TFD}}).
\end{align}
In the same way, we also obtain
\begin{align}
  S_{A_L A_R}^{(q)}(K_j\ket{\mathrm{TFD}})
  &=S_{B_L B_R}^{(q)}(K_j\ket{\mathrm{TFD}})\nonumber\\
  &\geq \sum_j r_{ij} S_{B_L B_R}^{(q)}(J_i K_j\ket{\mathrm{TFD}})\nonumber\\
  &= \sum_j r_{ij} S_{A_L A_R}^{(q)}(J_i K_j\ket{\mathrm{TFD}}),
\end{align}
where
\begin{align}
  &J_i = I_{A_L} \otimes I_{B_L} \otimes \ket{\overline{P_i}}\bra{\overline{P_i}}_{A_R} \otimes I_{B_R} ,\\
  &r_{ij} = \frac{\braket{\mathrm{TFD}|J_i K_j|\mathrm{TFD}}}{\braket{\mathrm{TFD}|K_j|\mathrm{TFD}}}.
\end{align}
Thus, we have
\begin{align}
  S_{A_L A_R}^{(q)}(\ket{\mathrm{TFD}})
  \geq \sum_{i, j} r_{ij} s_j S_{A_L A_R}^{(q)}(J_i K_j \ket{\mathrm{TFD}}).
\end{align}
Using the decomposition of $\ket{\mathrm{TFD}}$ shown in (6),  
\begin{align}
  J_i K_j \ket{\mathrm{TFD}} &= e^{-\beta H/2}\ket{P_{ij}}_L \otimes \ket{\overline{P_{ij}}}_R, \\
  r_{ij} s_j &= \frac{\braket{P_{ij}|e^{-\beta H}|P_{ij}}}{\sum_k\sum_l \braket{P_{kl}|e^{-\beta H}|P_{kl}}}\equiv p_{ij}, 
\end{align}
where $\ket{P_{ij}} \equiv \ket{P_i}_A \otimes \ket{P_j}_B $ .
Then, from the additivity and non-negativity of the R\'enyi entropy, we have
\begin{align}
  S_{A_L A_R}^{(q)}(\ket{\mathrm{TFD}})
  \geq& \sum_{i,j} p_{ij} S_{A_L A_R}^{(q)}(\ket{\mu_{ij}}_L \otimes \ket{\overline{P_{ij}}}_R) \nonumber\\
  =& \sum_{i,j} p_{ij} \left\{S_A^{(q)}(\ket{\mu_{ij}}) + S_A^{(q)}(\ket{\overline{P_{ij}}})\right\} \nonumber\\
  \geq& \sum_{i,j} p_{ij} S_A^{(q)}(\ket{\mu_{ij}}).
\end{align}    
Hence (7) is shown for $0<q\leq1$. 
\end{proof}

\section{Finite-size effects on the average entanglement entropy of METTS}

\begin{figure*}[!h]
    \centering
    \subfloat[Ising model ($\nu=x$)]{\label{fig:TFI_C0-L_qAxisX}\includegraphics[width=0.33\textwidth]{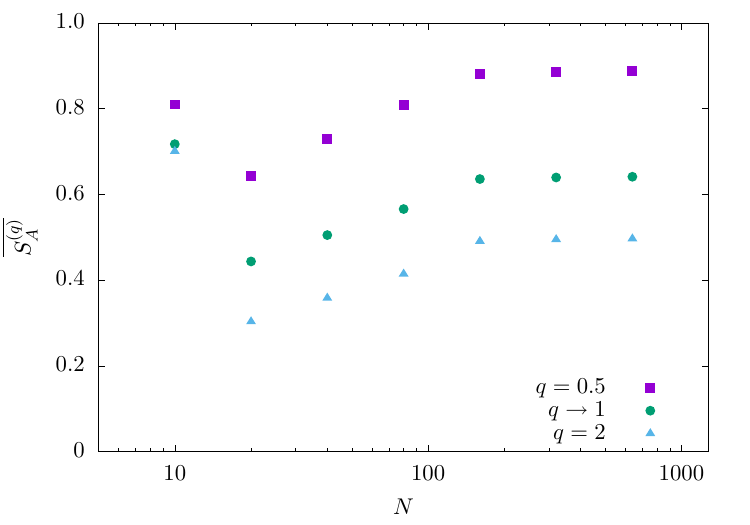}}\hfill
    \subfloat[Ising model ($\nu=z$)]{\label{fig:TFI_C0-L_qAxisZ}\includegraphics[width=0.33\textwidth]{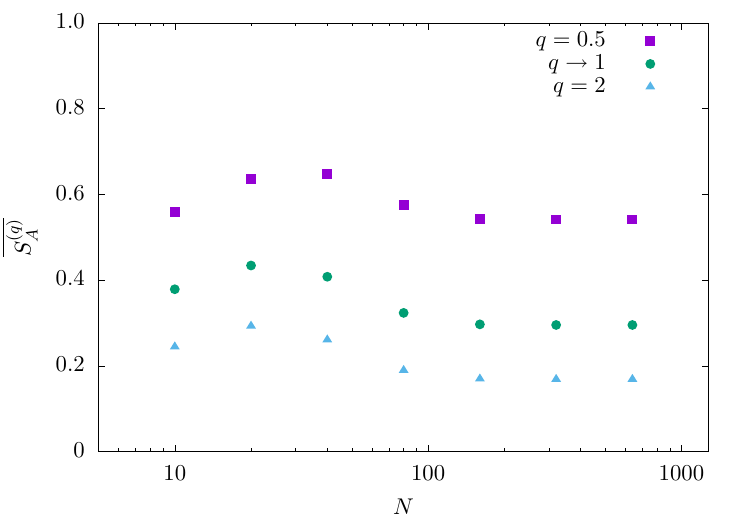}}\hfill
    \subfloat[Heisenberg model]{\label{fig:NNNHeis_C0-L}\includegraphics[width=0.33\textwidth]{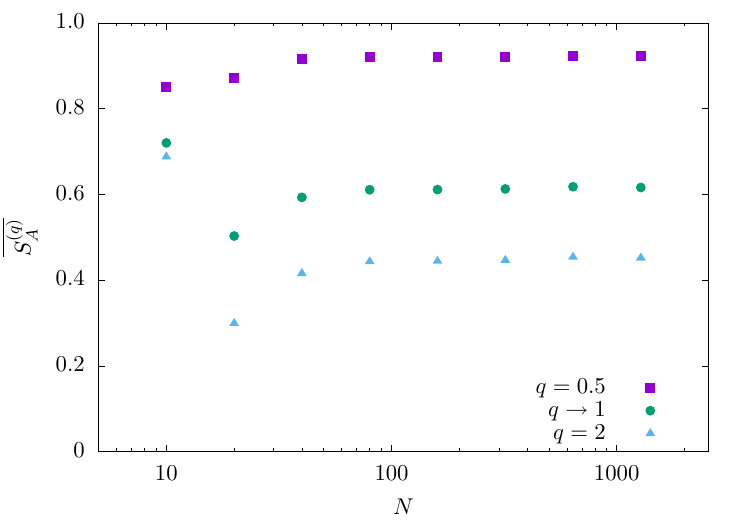}}
    \caption[]{$N$ dependence of the average entanglement entropy of METTS for the half-chain $A=\{1,2,\cdots,N/2\}$ in the critical transverse-field Ising model (a, b) and in the critical Heisenberg model with the next-nearest-neighbor interaction (c). Imaginary-time evolution is carried out using the second-order Trotter decomposition with time step $\delta\tau=0.04$ for the Ising model and $\delta\tau=0.01$ for the Heisenberg model. The data is average over $10000$ samples for the Ising model and $5000$ samples for the Heisenberg model.}
    \label{fig:C0-L}
\end{figure*}
In the main text, we perform numerical calculations on sufficiently large systems to neglect finite-size effects when investigating the inverse temperature dependence of the average entanglement entropy of METTS. Here, we examine the finite-size effects on the average entanglement entropy of METTS and justify the parameters adopted in our numerical calculations.

In the main text, numerical calculations are conducted for METTS at inverse temperatures ranging from $0 \leq \beta \leq \beta_\mathrm{max} = 128$ for the critical transverse-field Ising model and $0 \leq \beta \leq \beta_\mathrm{max} = 16$ for the critical Heisenberg model with next-nearest-neighbor interaction. Then, in Fig.~\ref{fig:C0-L}, we plot the system size dependence of the average entanglement entropy of METTS at the lowest temperature $\beta = \beta_\mathrm{max}$, where finite-size effects are most prominent within that range, for each model. We can confirm that setting $N=640$ for the critical transverse-field Ising model and $N=1280$ for the critical Heisenberg model with next-nearest-neighbor interaction is sufficient to ensure that finite-size effects can be adequately neglected.

\end{document}